\documentclass{iau}
\usepackage{graphicx}

\title[Maser measurements cross-matched with \textit{Gaia}] 
{Astrometric Galactic maser measurements cross-matched with \textit{Gaia}}
\author[Quiroga-Nu\~{n}ez et al.]   
{L.H.~Quiroga-Nu\~{n}ez$^{1,2}$
    H.J.~van~Langevelde$^{2,1}$
    M.J.~Reid$^{3}$
	L.O.~Sjouwerman$^{4}$
	Y.M.~Pihlstr\"{o}m$^{5}$
	A.G.A.~Brown$^{1}$    
 	\and J.A.~Green$^{6}$
}

\affiliation{$^1$ Leiden Observatory, Leiden University, \\ P.O. Box 9513, 2300 RA Leiden, The Netherlands. \\ email: {\tt quiroganunez@strw.leidenuniv.nl} \\[\affilskip]
$^2$Joint Institute for VLBI ERIC (JIVE), \\ Postbus 2, 7990 AA Dwingeloo, The Netherlands. \\[\affilskip]
$^3$Harvard-Smithsonian Center for Astrophysics, \\ 60 Garden Street, Cambridge, MA 02138, USA. \\ [\affilskip]
$^4$National Radio Astronomy Observatory, \\ P.O. Box 0, Lopezville Road 1001, Socorro, NM 87801, USA. \\ [\affilskip]
$^5$Department of Physics and Astronomy, University of New Mexico, \\ MSC07 4220, Albuquerque, NM 87131, USA. \\ [\affilskip]
$^6$CSIRO Astronomy and Space Science, Australia Telescope National Facility, \\ PO Box 76, Epping, NSW 1710, Australia. \\ [\affilskip]
}

\pubyear{2017}
\volume{334}  
\jname{Rediscovering our Galaxy}
\editors{C. Chiappini, I. Minchev, E. Starkenburg \& M. Valentini}
\begin{document}

\maketitle
\vspace*{-0.32 cm}
\begin{abstract}
Using the VLBA, the BeSSeL survey has provided distances and proper motions of young massive stars, allowing an accurate measure of the Galactic spiral structure. By the same technique, we are planning to map the inner Galaxy using positions and velocities of evolved stars (provided by the BAaDE survey). These radio astrometric measurements (BeSSeL and BAaDE) will be complementary to \textit{Gaia} results and the overlap will provide important clues on the intrinsic properties and population distribution of the stars in the bulge.
\end{abstract}

\vspace*{-0.38 cm}
\firstsection 
\section{Context and aim}

By monitoring the position of a maser bearing star in the sky with respect to a background extragalactic source, it is possible to estimate its proper motion and parallax (e.g., \cite{Honma08}). Maser astrometric measurements of high-mass star forming regions using VLBI have recently made substantial progress on our understanding of the Milky Way~(\cite{Reid14a}; \cite{Reid14b}). The Bar and Spiral Structure Legacy (BeSSeL) survey has obtained accurate positions, proper motions, line-of-sight velocities and distances to more than 100 young massive stars, which are associated with water and methanol maser emission. Using this 6D phase space information, the Galactocentric distance, solar rotation speed, rotation curve and solar motion have been refined~(\cite{Reid14b}). Moreover, simulations of the Galactic distribution of masers confirmed the accuracy of the parameter values found by BeSSeL (\cite{Quiroga17a}).

The Bulge Asymmetries and Dynamical Evolution (BAaDE) project surveys evolved stars with SiO maser emission at 43 and 86 GHz using the VLA and ALMA. BAaDE will eventually provide accurate positions and radial velocities for 20,000 stars mainly in the bulge (\cite{Sjouwerman16}). In principle, it is possible to estimate proper motions and parallaxes for some BAaDE targets following the technique used in BeSSeL. Moreover, since these radio astrometric measurements are made in the Galactic plane but are not affected by extinction, they can provide complementary information where optical surveys cannot reach. This will help to characterize the stellar population of the inner Galaxy, which is crucial to understand the Galactic dynamical evolution. 

\vspace*{-0.24cm}
\section{Cross-matching with \textit{Gaia} (TGAS) and future work}
\label{TGAS}

\begin{figure}[t!]
\vspace*{-0.25 cm}
\begin{center}\
\includegraphics[width=2.7in]{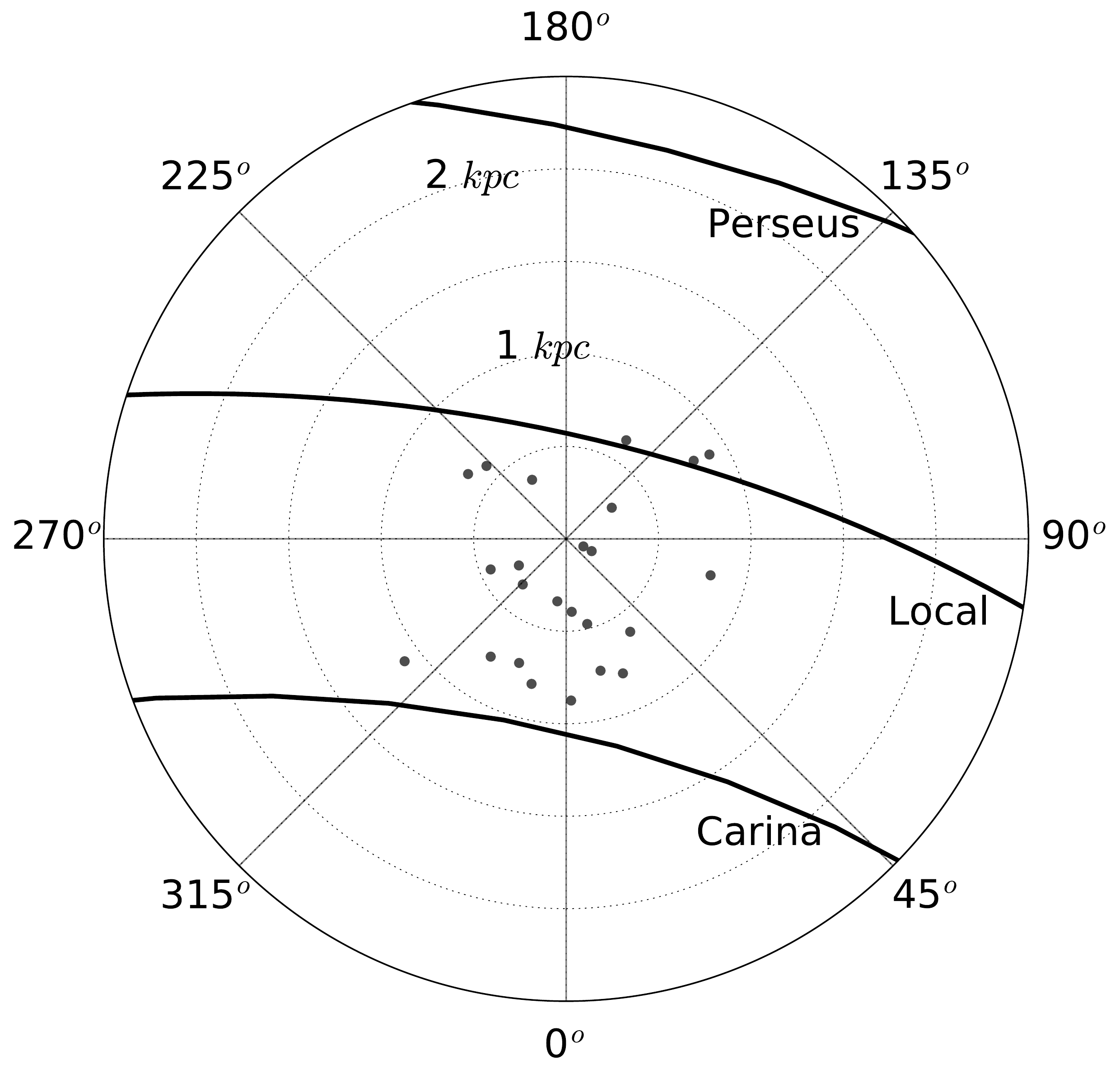} 
\includegraphics[width=2.4in]{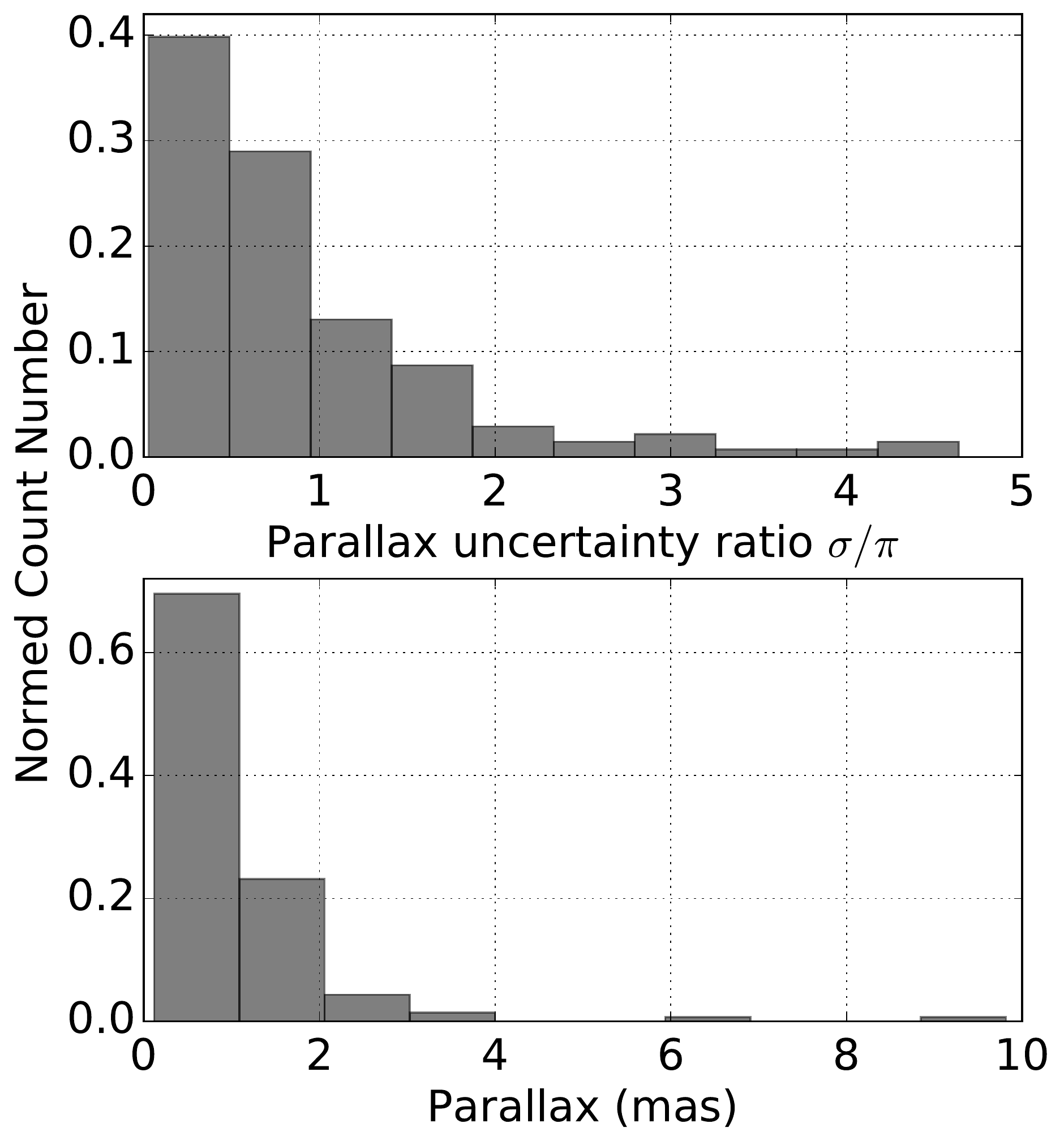} 
\vspace*{-0.25 cm}
\caption{\textbf{Left:} Distribution of the sample obtained by cross-matching BAaDE targets and TGAS over a top view of the Milky Way centered in the sun's position. Only the sources with $\sigma/\pi \le 0.3$ are plotted. \textbf{Right:} Distribution of the parallax uncertainty ratio and the parallax for the cross-match BAaDE-TGAS sample. Only the sources with $\sigma/\pi \le 5$ are plotted.}
\vspace*{-0.14cm}
\label{fig1}
\end{center}
\end{figure}

\cite{Quiroga17b} have cross-matched BAaDE targets with \textit{Gaia} DR1 resulting in more than 2,000 matches, where false positives were avoided by using several filters (distance analysis, IR color filters and variability). From that sample, 156 sources were observed with \textit{Hipparcos} and therefore are part of TGAS. The Galactic distribution of these sources is shown in the left panel of Fig.~\ref{fig1}, where only sources with low parallax uncertainties ($\sigma/\pi<0.3$) are shown. They represent 15$\%$ of the sample, and for those, we used the distance-parallax definition without prior assumption (\cite{Bailer15}). Right panel of Fig.~\ref{fig1} shows the distribution of the parallax uncertainty ratio ($\sigma/\pi$) and the parallax for those sources with $\sigma/\pi<5$. They represent 88$\%$ of the sample. As it can seen in Fig.~\ref{fig1}, the sample is confined to the foreground Galaxy.

We will soon have an unique sample of $\sim$2,000 evolved stars with radio (positions, radial velocities, SiO maser emission lines and rates from BAaDE), infrared ($J,H,K$ filters from 2MASS) and optical (parallaxes, proper motions, periods and photometry from \textit{Gaia}) bands to characterize the stellar population in the Galactic bulge. In addition, VLBI campaigns are being defined to provide astrometric information of AGB stars with bright masers and hence, some individual stars can be studied in detail. Finally, a fundamental comparison between radio and optical parallax-distance estimate can be performed.
\vspace*{-0.245 cm}
\begin{acknowledgements} 
This work has used data from the ESA mission \textit{Gaia}, processed by the \textit{Gaia} DPAC. This material is based upon work supported by the NSF under grant 1517970.
\end{acknowledgements}

\vspace*{-0.53 cm}
\end{document}